# Imprint from ferromagnetic skyrmions in an antiferromagnet via exchange bias

Kumari Gaurav Rana[1], Rafael Lopes Seeger[1], Sandra Ruiz-Gómez[2], Roméo Juge[1], Qiang Zhang[1], Kaushik Bairagi[1], Van Tuong Pham[1], Mohamed Belmeguenai[3], Stéphane Auffret[1], Michael Foerster[2], Lucia Aballe[2], Gilles Gaudin[1], Vincent Baltz [1,*], Olivier Boulle[1,**].

*1 Univ. Grenoble Alpes, CNRS, CEA, Grenoble INP\*\*\*, IRIG-SPINTEC, F-38000 Grenoble, France*

*2 Alba Synchrotron Light Facility, Carrer de la llum 2-26, 08290, Cerdanyola del Valles, Barcelona, Spain*

*3 Laboratoire des Sciences des Procédés et des Matériaux, CNRS, Université Paris 13, 93430 Villetaneuse, France*

\* vincent.baltz@cea.fr\*\* olivier.boulle@cea.fr

**Abstract**

Magnetic skyrmions are topological spin textures holding great potential as nanoscale information carriers. Recently, skyrmions have been predicted in antiferromagnets, with key advantages in terms of stability, size, and dynamical properties over their ferromagnetic analogs. However, their experimental demonstration is still lacking. Here, we show the imprint from ferromagnetic skyrmions into a thin film of an IrMn antiferromagnet, at room temperature and zero external magnetic field, using exchange-bias. Using high-spatial-resolution x-ray magnetic circular dichroism photoemission electron microscopy (XMCD-PEEM), we observed the imprinted spin textures within the IrMn from the XMCD signal of the uncompensated Mn spins at the interface with the ferromagnet. This result opens up a path for logic and memory devices based on skyrmion manipulation in antiferromagnets.

Magnetic textures such as skyrmions are currently attracting a lot of attention due to their rich spin physics and their high potential for storage and logic computing. [1,2] Skyrmions are topological magnetic defects in the magnetic spin structure. [3,4] They can be found either in the bulk or interfacial systems lacking inversion symmetry, and require Dzyaloshinskii-Moriya interaction (DMI). [5–7] They hold great promise for future spintronic applications as they can be as small as few nanometers, and can be created, manipulated and annihilated electrically, in particular by an electron or a magnon flow [8,9].

Since antiferromagnets [10,11] can be magnetic at the atomic scale and non-magnetic at the macroscopic scale, skyrmions in antiferromagnets, [12,13] also known as antiferromagnetic skyrmions, may present several advantages over their ferromagnetic analogs. More specifically, they combine key features for applications in the field of spintronics as: (i) they produce no dipolar fields, making them stable at the nanometer scale in zero external magnetic field; (ii) they are robust against perturbation due to magnetic fields, which is beneficial for data retention; (iii) they exhibit zero net topological charge, [12,14] thus eliminating the unwanted transverse velocity related to the skyrmion Hall effect, [15,16] thereby ensuring a straight skyrmion trajectory with enhanced mobility. [12] However, since they lack net magnetization the experimental observation and nucleation of antiferromagnetic skyrmions is challenging. Nevertheless, the direct observation of antiferromagnetic spin textures can be achieved, [17] but in large-scale facilities with element-sensitive techniques like X-ray absorption spectroscopy, [18–20] specific local probe techniques such as spin-polarized scanning tunneling microscopy [21,22] and quantum sensing with single spins (nitrogen vacancies) in diamond, [23–25] or optical second harmonic generation [26] and thermal gradient microscopy [27]. Antiferromagnetic domains and domain walls in NiO, [18,28] $BiFeO_3$, [23,26] $Cr_2O_3$, [24] CuMnAs [25] and $Mn_3Sn$ [27], and vortex states in IrMn [19] and NiO [20] layers were for example investigated in those ways. Recently, fractional antiferromagnetic skyrmion lattices in bulk $MnSc_2S_4$ were also observed at cryogenic temperature using neutron scattering experiments [29] as well as antiferromagnetic half-skyrmions and bimerons in $\alpha$-$Fe_2O_3$ at room temperature [30], where the transition over magnetic phases using low temperature cycling was used to nucleate the spin textures. An important bottleneck is the nucleation of skyrmions in thin films of antiferromagnets at room temperature. Application of an external magnetic field, like in ferromagnetic [4] is ineffective for actual antiferromagnets. We note that, for the case of synthetic antiferromagnets [31], the low antiferromagnetic coupling between two ferromagnetic skyrmions makes these skyrmions sensitive to external magnetic field. An antiferromagnet is also more robust against dynamical torque deformation, since its spin structure relies on strong interatomic exchange interactions.

An alternative way to manipulate the order parameter of an antiferromagnet, is to take advantage of the strong exchange bias interaction between the antiferromagnet and an adjacent ferromagnet [32] to imprint a ferromagnetic configuration in the antiferromagnet.[11]. In a first step, exchange bias interaction is quenched by raising the sample temperature above the blocking temperature ($T_B$) of the ferromagnet/antiferromagnet bilayer. The antiferromagnetic layer loses its ability to pin the magnetization of the adjacent ferromagnet. The latter can then be considered as a ferromagnetic single layer in which it is possible to nucleate different types of spin textures by conventional means. In a second step, the bilayer is cooled below $T_B$ causing the moments in the antiferromagnet to align with those of the ferromagnet due to the exchange bias coupling. This procedure was used to imprint multidomain states and magnetic textures in antiferromagnets, by preparing the ferromagnet above $T_B$ in specific magnetic configurations, *e.g.* multidomain states by demagnetization under decreasing, alternating magnetic field, [33,34] or vortex states by using magnetic nanostructures. [19,20]

Based on this approach, we show the imprint from ferromagnetic skyrmions into a thin film of an IrMn antiferromagnet, at room temperature and zero external magnetic field, using exchange-bias. Using the high spatial resolution magnetic microscopy technique XMCD-PEEM, we observe the imprinted spin textures within the IrMn layer from the XMCD signal of the uncompensated Mn spins.

The sample investigated consisted of a //Ta(3)/Cu(3)/IrMn(5)/Pt(0.5)/Co(0.3)/NiFe(0.87)/Al(2) (nm) multilayer, from substrate to surface. The polycrystalline stack was deposited at room temperature by magnetron sputtering on a Si/SiO$_2$(500) wafer at a pressure of 2.3 x 10$^{-3}$ mbar under argon. The IrMn antiferromagnet was deposited from an Ir$_{20}$Mn$_{80}$ (at. %) target and the NiFe ferromagnet was deposited from a Ni$_{81}$Fe$_{19}$ (at. %) permalloy target. A Ta(3)/Cu(3) seed bilayer was used to promote the growth of the antiferromagnetic (111)-textured fcc phase of the IrMn alloy. The composition of the stack was carefully optimized in order to stabilize skyrmions in the Co/NiFe layer and allow for their observations using X-ray magnetic microscopy. The ultra-thin intermediate bilayer of Pt(0.5)/Co(0.3) allows us to achieve large perpendicular magnetic anisotropy as well as a large interfacial DMI [4], [35] without magnetically decoupling the IrMn and NiFe layers. This thickness of the NiFe layer was chosen in the vicinity of the planar-to-perpendicular anisotropy transition of the ferromagnet, therefore allowing us to promote the formation of skyrmions. [4] The choice of the stacking order for the IrMn and NiFe layers, combined with the reduced thickness of the NiFe layer, was such that the sensitivity of photoemission electron microscopy to Mn spins at the IrMn surface was most favorable. A 2-nm-thick Al cap, a light element, was finally deposited to form a protective and transparent Al(2)O$_x$ film after oxidation in air.

To achieve exchange bias, the sample temperature was first raised to 250°C, above $T_B$, [11] kept for 30 minutes and cooled to room temperature in an external magnetic field applied along the out-of-plane direction. The amplitude of the external field was of 0.57 T, *i.e.* sufficient to saturate the NiFe ferromagnetic layer. Then, the sample temperature was raised a second time to 250°C for 30 minutes at zero field and cooled to room temperature [36]. The out-of-plane magnetic-field-dependence of the Kerr signal as measured at room temperature (Fig. 1) is shifted with respect to zero-field by an exchange bias field, of around 50 mT. The linear and anhysteretic reversal of the signal further indicates stripe domain reversal.

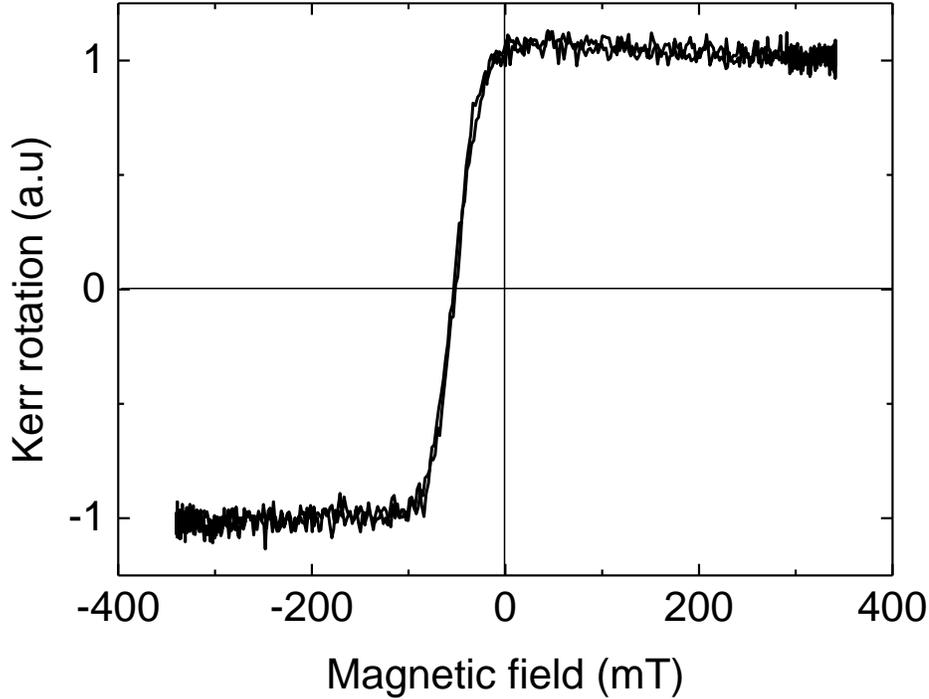

**Figure 1** Out-of-plane magnetic-field-dependence of the Kerr signal as measured at room temperature for a //Ta(3)/Cu(3)/IrMn(5)/Pt(0.5)/Co(0.3)/NiFe(0.7)/Al(2) (nm) stack after the annealing procedure.

Brillouin Light Scattering experiments were carried out to extract the amplitude of the DMI which leads to D = −0.30 mJ/m$^2$ [36]. The negative value for D favors a left-handed chirality, consistently with [4,35]. The critical D associated with the Bloch to Néel domain wall transition writes [37] $D_c = 4\mu_0 M_s^2 t \ln2/2\pi^2$ with t the film thickness . It can be estimated to be $D_c$=0.16 mJ/m² in our stack. Thus, D>$D_c$ and the domain wall and skyrmions in the Co/NiFe ferromagnet are expected to be of the chiral left-handed Néel type.

To gain insight into the imprint of spin textures at the IrMn layer interface through exchange bias, element-specific X-ray magnetic circular dichroism photoemission electron microscopy (XMCD-PEEM) experiments were subsequently carried out at room temperature and in zero external magnetic field. These experiments were performed on the SPELEEM III microscope (Elimtec GmbH) at the CIRCE beamline in the ALBA synchrotron. [38] Typical X-ray absorption spectra (XAS) integrated over an area of about 300 µm$^2$ are given in Fig. 2(a) (circular polarization) and 2(b) (linear polarization). The X-ray magnetic circular dichroism (XMCD) in Fig 2(a) shows the existence of a small net magnetic moment in IrMn, ascribed to the frozen uncompensated spins after the cooling procedure. The corresponding X-ray magnetic linear dichroism (XMLD) spectra (Fig. 2(b)) indicates that the overall orientation of the Mn spins is tilted out-of-plane, [39] in agreement with what is expected from the cooling procedure.

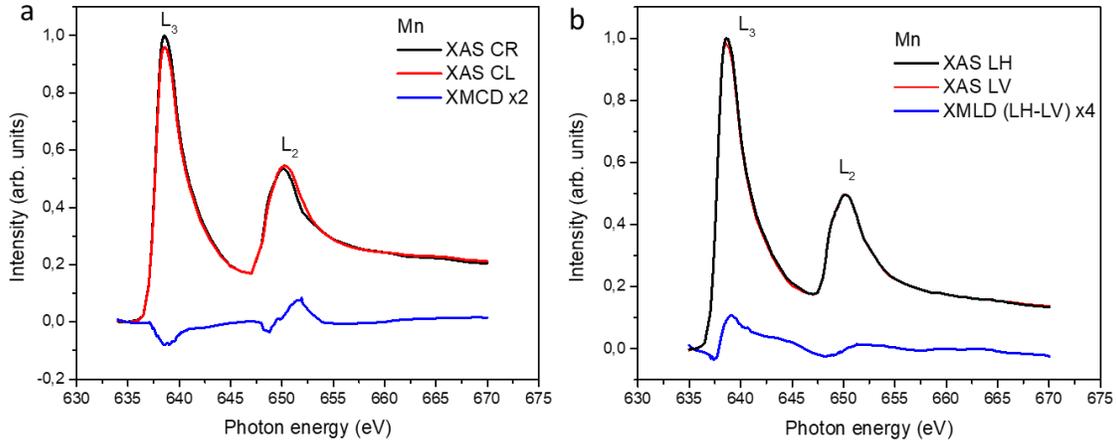

**Figure 2 Energy absorption spectra corresponding to X-ray magnetic (a) circular and (b) linear dichroism photoemission electron experiments for a //Ta(3)/Cu(3)/IrMn(5)/Pt(0.5)/Co(0.3)/NiFe(0.87)/Al(2) (nm) stack. Two polarities are shown. In (a), CR and CL stands for circular right and circular left polarization respectively. XMCD is the difference of the CL and CR signal; its amplitude is multiplied by a factor 2. In (b), LH and LV stands for linear horizontal and linear vertical polarization. The energy window is focused in the vicinity of the L-edges of the Mn element.**

Spatially resolved XMCD-PEEM images were recorded at the Fe, and Mn L-edges for right- and left-circularly polarized X-rays. The resulting magnetic contrast image at the Fe edge (Fig. 3(a)) indicates that magnetic skyrmions are stabilized in zero external magnetic field in the NiFe layer. Images at the Mn edge (Fig. 3(b)) provide information on the non-compensated Mn spins at the top interface of the IrMn layer. For some regions, spin textures can clearly be observed, whose shape and position coincide with the skyrmions in the NiFe ferromagnet (Fig. 3(c)). These results demonstrate that the skyrmionic spin texture in the NiFe ferromagnet is replicated in the interfacial Mn spins of the IrMn antiferromagnet. Although we cannot conclude on the penetration depth of the spin texture due to the complex spin structure of the IrMn antiferromagnet [11] combined with the weighted depth-sensitivity of the measurement, we note that earlier works have shown that exchange bias made it possible to imprint other spin textures in the depth of the antiferromagnet, down to at least 3 nm for vortices in CoO and NiO, [20] and exchange springs in IrMn [40] layers.

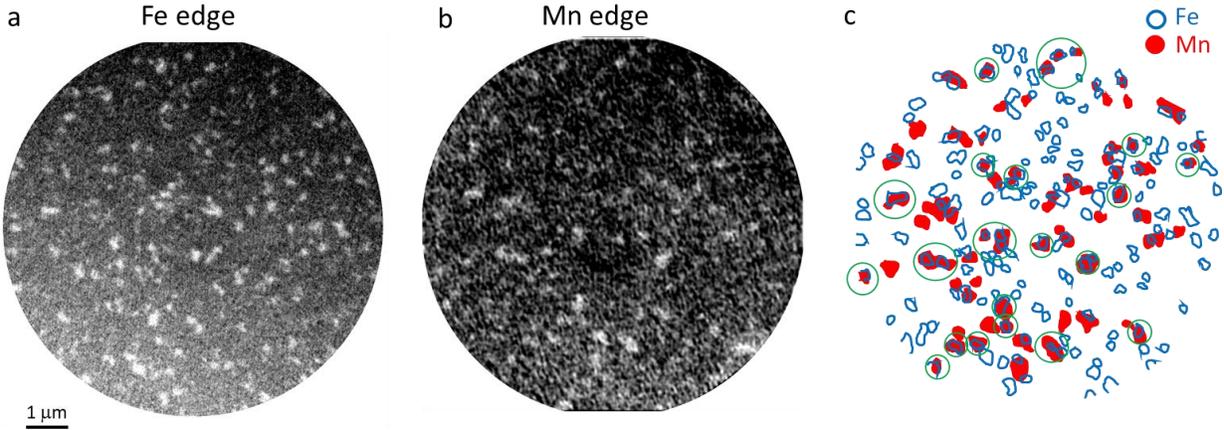

**Figure 3** Images corresponding to the NiFe and IrMn magnetic contrasts for a //Ta(3)/Cu(3)/IrMn(5)/Pt(0.5)/Co(0.3)/NiFe(0.87)/Al(2) (nm) stack. The images are obtained at room temperature and in zero external magnetic field, by XMCD-PEEM at the L-edges of the energy absorption spectra of (a) the Fe and (b) the Mn elements, respectively. (c) Superposition of the contours of textures for the Fe (open, blue) and for the Mn (filled, red) elements. The green circles indicate area where an overlap between Fe and Mn features is observed.

From Fig. 3(c), we can also observe that conformity between the ferromagnetic and antiferromagnetic spin textures does not hold everywhere. For some areas, skyrmions in the NiFe layer are not replicated in the IrMn layer, and for other areas, textures in the IrMn layer are replicated but are no more facing textures in the NiFe layer. We found that 43 features at the Mn-edge overlap with the 133 features at the Fe-edge. These results can be explained on the one hand by the limited sensitivity at the Mn-edge resulting from the small XMCD signal. This gives a lower bound overlap as we cannot exclude that some features at the Mn-edge are not detected. On the other hand, these results can be accounted for the known spatial distribution of blocking temperature ($\Delta T_B$) [32,36,41]. More specifically, for areas where $T_B \geq 300$ K, the magnetic configuration of the NiFe ferromagnet can be stabilized in the 5 nm-thick IrMn antiferromagnet at 300 K by a cooling procedure from $T_B$ to 300 K, in contrast to areas where $T_B < 300$ K, in which case, textures cannot be stabilized in this way in the IrMn antiferromagnet [42,43], [36]. Analysis of the blocking temperature distribution ([36], Fig. S4) returned that 57 % of the antiferromagnet area is exchange coupled to the ferromagnet at room temperature, setting the upper bound for overlap. Note that there has been so far only indirect insights on such spatial distributions, for example through the correlation between disordered magnetic phases spread above ferromagnetic/antiferromagnetic thin films and device-to-device variability of exchange bias in spintronic applications, after patterning the thin film [36,44]. Here, our observation would provide a direct observation of the spatial distribution of blocking temperature in an exchange-biased stack. Additional experiments also show that the spin textures in the IrMn are affected by changes in the domains in the NiFe layer: the non-compensated spins in the antiferromagnet are dragged by the spins in the ferromagnet upon application of an external field [36,45]. Also, other experiments show that the second annealing, followed by zero field cooling, is not required to replicate the skyrmions in the IrMn [36]. This underlines that exchange bias plays a significant role in the stabilization of the spin texture in the IrMn.

In conclusion, we demonstrate the imprint from ferromagnetic skyrmions into a sputtered IrMn antiferromagnetic thin film, at room temperature and in zero external magnetic field, using exchange bias. Element-specific X-ray magnetic microscopy allows the direct observation of the imprinted spin texture at the interface of the antiferromagnet, from the uncompensated Mn spins at the interface. This study paves the way for future advances, since several spin-dependent transport effects in antiferromagnetic skyrmions, like topological Hall [46] or spin-orbit torque [47] effects, have been theoretically predicted and have yet to be experimentally demonstrated, along with the closely related promising applications opening up a path for logic and memory devices based on skyrmion manipulation in antiferromagnets.

## Supplementary Materials

See supplementary materials for additional data on the determination of the Dzyaloshinskii-Moriya interaction using Brillouin light scattering experiments, effect of external magnetic fields and annealing procedure on the spin textures, as well as the blocking temperature distribution.

## Acknowledgement

We acknowledge financial support from the French national research agency (ANR) [Grant Numbers ANR-15-CE24-0015-01 and ANR-17-CE24-0045], , the bottom-up exploratory program of the CEA (Grant Number PE-18P31-ELSA), the Alliance Hubert Curien programme (PHC) [Grant Number 46298XC], and the American defense advanced research project agency (DARPA) TEE program [Grant Number MIPR HR0011831554].

## Data availability

The data that support the findings of this study are available from the corresponding author upon reasonable request.